\documentclass[aps,prc,showpacs,nofootinbib,twocolumn]{revtex4-1} 
\usepackage{graphicx}
\begin{document}
\title{$\Sigma^-$ admixture in the $^{10}_\Lambda$Li hypernucleus with a repulsive $\Sigma$-nucleus potential}
\author{D.\ E.\ Lanskoy}
\email{lanskoy@sinp.msu.ru}
\author{A.\ Sinyakova}
\email{arina.sinyakova@gmail.com}
 \affiliation{Skobeltsyn Institute of Nuclear
Physics, Lomonosov Moscow State University, 119991 Moscow, Russia}
\date{\today}

\begin{abstract}
$\Sigma^-$ hyperon component of the $^{10}_\Lambda$Li wave function is studied. The $\Sigma^-$ admixture is vital for production of neutron-rich $\Lambda$ hypernuclei via mesonic beams. We use a simplified shell model wave function for the $\Lambda$ channel and calculate the $\Sigma^-$ admixture directly from coupled equations. Probability of the $\Sigma^-$ admixture for realistic repulsive $\Sigma^-$-nucleus potentials is less by several times than that for attractive potentials and does not exceed 0.1\%. We conclude that the cross sections of the $^{10}\mbox{B}(\pi^-,K^+){}^{10}_\Lambda\mbox{Li}$ reaction measured at KEK cannot be explained by production via $\Sigma^-$ admixture as a doorway state.\end{abstract}
\maketitle

\section{\label{s1}Introduction}
Neutron-rich $\Lambda$ hypernuclei were discussed by theorists numerously (see, e.g., \cite{Dal,Maj,TL99,Akk,Hiy,GaMi}). Some data on $^{7}_\Lambda$He and $^{8}_\Lambda$He have been obtained from the old emulsion experiments \cite{DP}. A new stage of study of neutron-rich hypernuclei was opened by the KEK experiment \cite{Saha}. The $^{10}_\Lambda$Li hypernucleus was produced by the $^{10}\mbox{B}(\pi^-,K^+)$ reaction. Because of poor statistics, it was impossible to measure the binding energy of $^{10}_\Lambda$Li, but the production cross section has been determined. Later, few events interpreted as production of $^{6}_\Lambda$H were reported by the FINUDA collaboration \cite{FINH6}. They studied the $^{6}\mbox{Li}(K^-,\pi^+)$ reaction at rest and detected pions from weak decay of the final hypernucleus. Hypernucleus $^{7}_\Lambda$He was observed also by the $(e,eK^+)$ reaction \cite{CEB}. In addition, several experiments \cite{Kub,FINa,FINHe9,JPRC} are known, which established only upper bounds for production probabilities.

Production of neutron-rich $\Lambda$ hypernuclei remains to be a difficult task not only because typical cross sections are small. The production dynamics is  poorly understood too, and theoretical predictions for production probabilities are problematic. The most evident mechanism of the $(\pi^-,K^+)$ and $(K^-,\pi^+)$ reactions is the two-step one with meson charge exchange, for instance, $\pi^-p\to \pi^0n$ followed by $\pi^0p\to K^+\Lambda$. Another mechanism was considered in \cite{TL01}, namely, one-step process $\pi^-p\to K^+\Sigma^-$. A small $\Sigma^-$ admixture in a $\Lambda$ hypernucleus appears from the $\Lambda n-\Sigma^-p$ interaction and serves as a doorway state. The production amplitudes are small due to two-step nature of the first mechanism and smallness of the $\Sigma^-$ admixture providing the second mechanism. This is not obvious which mechanism is more productive. Note also that the third mechanism, production via $\Delta$ admixture in the initial nucleus \cite{L04}, remains unexplored.

At first, it was suggested that the two-step mechanism is dominant for the $(\pi^-,K^+)$ process, and the two-step cross section for the $^{10}\mbox{B}(\pi^-,K^+){}^{10}_\Lambda\mbox{Li}$ reaction was estimated \cite{TL03,L04}. Experiment \cite{Saha} gave substantially (by several times) lower cross sections. This is not surprising in view of many poorly known input parameters involved in the calculation. More important, the experiment showed that the production cross section is higher at incident pion momentum $k_\pi=1.2$ GeV/c than at $k_\pi=1.05$ GeV/c while the theory predicted the opposite $k_\pi$ dependence. Thus the measured cross sections seem to be incompatible with the assumption of the two-step process dominance.

Harada \textit{et al.} \cite{HUH} suggested that, otherwise, the one-step process is prevailing. Using the distorted wave impulse approximation and the Green function method for the reaction amplitude, they succeeded to reproduce the observed $k_\pi$ dependence of the cross section. Its absolute value agrees with the experiment if probability of $\Sigma^-$ admixture $p_{\Sigma^-}$ is as large as $0.5\div 0.7$\% \cite{HUH}.

Probability $p_{\Sigma^-}$ is the key quantity in this problem. We do not discuss here a long history of $\Sigma$ admixture studies concerning various topics for various $\Lambda$ hypernuclei, see, e.g., \cite{DvH,Bod,Gal,Shin,Aka,Aka2}. As for specific $^{10}_\Lambda$Li, the $\Lambda N-\Sigma N$ mixing has been treated in the framework of the shell model \cite{Har1,Har2,Mil,GaMi}. We apply here another approach which is free from some restrictions inherent in the previous studies.

In Sec.\ \ref{s2}, we introduce our model and compare it with other approaches. Details of the calculation and results are presented in Sec.\ \ref{s3}. In Sec.\ \ref{s4}, we discuss implications of our results and give some concluding remarks.

\section{\label{s2}Treatment of $\Lambda n-\Sigma^-p$ mixing in $\Lambda$ hypernuclei}

Considering the $\Lambda N-\Sigma N$ coupling in $\Lambda$ hypernuclei, one comes to Hamiltonian
\begin{equation}
H=\left(\begin{array}{cc}H_{\Lambda\Lambda}&V_{\Lambda\Sigma}\\V_{\Lambda\Sigma}&H_{\Sigma\Sigma}\end{array}\right).\label{e1}
\end{equation}
Here $H_{\Lambda\Lambda}$ contains a $\Lambda$-nucleus potential, which is, strictly speaking, different from a phenomenological potential used in a single-channel picture. The latter one involves already a contribution arising from the $\Lambda N-\Sigma N$ coupling. The $\Sigma$-nucleus potential in $H_{\Sigma\Sigma}$ is usually assumed to be the same as its real part for a real $\Sigma$ hyperon, though a $\Sigma$ hyperon in $\Lambda$ hypernuclei is substantially virtual.
The imaginary part of the $\Sigma$-nucleus potential, describing the $\Sigma N\to\Lambda N$ conversion of a real $\Sigma$ hyperon, is irrelevant to this problem. But the $\Lambda N-\Sigma N$ interaction determines nondiagonal term $V_{\Lambda\Sigma}$.

In the shell model approaches \cite{Har1,Har2,Mil,GaMi}, configurations with $\Lambda$ and $\Sigma$ hyperons are treated at the same footing. This means that eigenstates of $H_{\Lambda\Lambda}$ and $H_{\Sigma\Sigma}$ are mixed by nondiagonal interaction $V_{\Lambda\Sigma}$ just like various configurations in the $\Lambda$ sector are mixed by a residual interaction. The shell model is famous in nuclear and hypernuclear physics and has been applied successfully to spectra of $p$ shell $\Lambda$ hypernuclei \cite{GSD,Fet,MilCh}.

However, being extended to the $\Sigma$ sector, this approach encounters some problems. Since $M_\Sigma-M_\Lambda$ is about 80 MeV, all eigenvalues of $H_{\Sigma\Sigma}$ lie rather highly and far from energies of the corresponding $\Lambda$ hypernucleus. Therefore, differences between eigenenergies of $H_{\Sigma\Sigma}$ and the actual hypernuclear energy (denominators in a perturbative expansion) have comparable values for many
$H_{\Sigma\Sigma}$ eigenstates including continuum ones. Thus, it is difficult to truncate properly the perturbative expansion in the $H_{\Sigma\Sigma}$ eigenstates. Moreover, usually only bound eigenstates are taken into account in the shell model. Even if the $\Sigma$-nucleus potential is attractive, only the $1s$ hyperon state is typically bound in light hypernuclei\footnote{Strictly speaking, the $H_{\Sigma\Sigma}$ spectrum for $\Sigma^-$ contains also the multitude of Coulombic bound states. However, being rather spatially extended, Coulombic states unlikely give a sizable contribution to the expansion at least in light systems.}. So all other (continuum) eigenstates lying not so far from the bound state are simply dropped. Strictly following these lines, one may conclude that the mixing is completely impossible for a repulsive (or even slightly attractive) $\Sigma$-nucleus potential, which is evidently wrong. Note that after the KEK experiment \cite{Sah1}, interaction of a $\Sigma^-$ hyperon with a neutron-rich nucleus is believed to be repulsive.

We apply here another approach. The model is rather simple, but does not include explicitly any expansion in eigenstates of $H_{\Sigma\Sigma}$, which seems to be an inappropriate basis in this problem. Therefore, we deal with repulsive and attractive $\Sigma$-nucleus potentials with equal ease.

First, let us write down the $^{10}_\Lambda$Li wave function as
\begin{equation}
\begin{array}{lll}
\left|{}^{10}_\Lambda\mbox{Li}, J^\pi\right>&=&a\left|{}^9\mbox{Li}\otimes\Lambda (n_\Lambda l_\Lambda j_\Lambda),J^\pi\right>\\ &&+b\left|{}^9\mbox{Be}\otimes\Sigma^-(l_\Sigma j_\Sigma), J^\pi\right>.\label{e2}
\end{array}
\end{equation}
We consider only the ground state doublet of $^{10}_\Lambda$Li, i.e., the $\Lambda$ hyperon in the $1s_{1/2}$ state moving in the field of $^9$Li(gs). Since spin and parity of $^9$Li(gs) are $3/2^-$, the lowest levels of $^{10}_\Lambda$Li are $1^-$ and $2^-$. We restrict our consideration to an $s$ state of $\Sigma^-$ hyperon and a $3/2^-$ state of $^9$Be.
 
Since one neutron is involved in the $\Lambda n-\Sigma^- p$ transition, we use the technique of coefficients of fractional parentage:
\begin{equation}
\left|{}^9\mbox{Li}\right>=\sum_{i=1}^k \alpha_i\left|\left({}^8\mbox{Li}\otimes n\right)_i, 3/2^-\right>.\label{e3}
\end{equation}
We take into account three lowest states of $^8$Li: 2$^+$ (gs), 1$^+$ (0.98 MeV), and 3$^+$ (2.26 MeV).

Applying Hamiltonian (\ref{e1}) to wave function (\ref{e2}), one gets a system of $2k$ coupled equations. However, two-particle interaction $V_{\Lambda\Sigma}$ does not mix different states of $^8$Li, which is an inert core in this problem. Therefore, the system separates to $k$ independent pairs of equations. For the $i$th component, we have
\begin{equation}
\begin{array}{rcrcl}
H_{\Lambda\Lambda}\psi^i_{\Lambda}&+&V_{\Lambda\Sigma}\psi^i_{\Sigma}&=&e_{\Lambda}\psi^i_{\Lambda}\\
V_{\Lambda\Sigma}\psi^i_{\Lambda}&+&H_{\Sigma\Sigma}\psi^i_{\Sigma}&=&e_{\Sigma}\psi^i_{\Sigma}.\label{e4}
\end{array}
\end{equation} 
Here $e_\Sigma=e_\Lambda+B({}^9\mbox{Be})-B({}^9\mbox{Li})+M_\Lambda-M_\Sigma+M_n-M_p$.

We stress that we do not try to solve the general problem of the $\Lambda N-\Sigma N$ coupling in $^{10}_\Lambda$Li, including, for instance, corresponding energy shifts. Our goal is only to find the specific $\Sigma^-$ component, which is of vital importance for the production reactions. That is why we do not insert explicitly other $\Sigma$ channels (as well as many various channels in the $\Lambda$ sector, which may be comparable in probabilities with $\Sigma$ channels) to the equations. It is inferred that all other channels are included effectively into $H_{\Lambda\Lambda}$. Therefore, $H_{\Lambda\Lambda}$ should contain just a phenomenological $\Lambda$-nucleus potential in this approach. The single $\Sigma^-$ channel treated explicitly is small and cannot modify the $\Lambda$-nucleus interaction significantly. 

Since the $\Sigma^-$ admixture is small, it cannot also modify meaningfully the wave function in the $\Lambda$ channel. Therefore, we may treat system (\ref{e4}) perturbatively. First, $\psi_\Lambda^i$ is found from the first equation without the coupling term. Then, the second equation is solved for $\psi^i_\Sigma$. For radial wave function $u_\Sigma^i$, the equation is as follows:
\begin{equation}
\left(-\frac{1}{2\mu_{\Sigma}}\frac{d^2}{dr^2}+
V_{\Sigma\Sigma}(r)-e_{\Sigma}
\right)u_{\Sigma}^i(r)=S^i(r),\label{e5}
\end{equation}
where source term $S^i(r)$ arises from the matrix element of the $\Lambda n\leftrightarrow\Sigma^-p$ transition. Finally, the $\Sigma^-$ admixture wave function is 
\begin{equation}
u_\Sigma(r)=\sum^k_{i=1}u^i_\Sigma(r)\label{e6}
\end{equation}
and the probability
\begin{equation}
p_{\Sigma^-}=\int^\infty_0u^2_\Sigma(r)dr.\label{e7}
\end{equation}

The resulting wave function cannot be recognized as the $1s$ one since it does not correspond to any eigenstate of $H_{\Sigma\Sigma}$. In terms of these eigenstates, it is generally a superposition of bound and continuum $s$ eigenstates. 

The $3/2^-$ state of $^9$Be in (\ref{e2}) is also not the ground state of $^9$Be, but a superposition of different $^9\mbox{Be}(3/2^-)$ states. Coefficients of fractional parentage $^9\mbox{Be}\to{}^8\mbox{Li}+p$ are determined by the coefficients of expansion (\ref{e3}) and the coupling. It should be noted here that we neglect $^9$Be states with other spins and parities, which is a restriction of the model.

The $\Lambda n-\Sigma^-p$ interaction is chosen as a central spin-dependent potential
\begin{equation}
V_{\Lambda\Sigma}({\bf r}_1-{\bf r}_2)=V_0({\bf r}_1-{\bf r}_2)+V_\sigma({\bf r}_1-{\bf r}_2)\mbox{\boldmath$\sigma_1\cdot\sigma_2$}.\label{e8}
\end{equation}
It is seen from \cite{GaMi} that the tensor coupling interaction plays typically a minor role in $p$ shell $\Lambda$ hypernuclei so we neglect the tensor potential in (\ref{e8}).

The approach described above is similar to that used in \cite{TL01} for $^{12}_\Lambda$Be and $^{16}_\Lambda$C. However, only one term in (\ref{e3}) corresponding to the ground state of the inert core was taken into account in that paper.

\section{\label{s3}Results and discussion}

The most important input in our calculation is, of course, coupling potential $V_{\Lambda\Sigma}$. Empirical knowledge of this interaction is nearly absent, and various theoretical models give rather different predictions. We do not make here any selection of known potentials. Instead, we examine interactions available in the literature and try to establish limits, in which the true results may be confined. We use the effective interactions of the so-called YNG type (quoted in \cite{Kohno}) derived by Yamamoto from the old Nijmegen models: model D with a hard core (NHCD) \cite{NHCD} and the early soft-core model (NSC89) \cite{NSC89} as well as much more recent interactions ESC08a, ESC08a$^{\prime\prime}$, and ESC08b (Nijmegen extended soft-core models) \cite{ESC}. We ignore the tensor parts of the ESC interactions. All these potentials were obtained as effective interactions by the Brueckner $G$ matrix calculation (for details, see \cite{Yam94,ESC}). We use also potential D2 from \cite{Aka}.

Another input is coefficients of fractional parentage $^9\mbox{Li}\to{}^8\mbox{Li}+n$. We consider the corresponding spectroscopic factors extracted from experiments and calculated theoretically \cite{Kan,Jep,Kwas,Tim}. The spectroscopic factors $S_i$ are anticipated to obey the known sum rule $\sigma=\sum_i S_i=N$, where the sum is over all $^8$Li states and $N=4$ is the number of neutrons in the $p$ shell. However, the sum over the three $^8$Li states ($2^+$, $1^+$, and $3^+$) for the considered sets of the spectroscopic factors substantially deviates from 4. Usually, $\sigma<4$, which can be explained by a missed contribution of higher $^8$Li states. Spectroscopic factors from \cite{Jep} give, however, $\sigma>4$. The results obtained with various sets of the spectroscopic factors therefore differ from each other drastically. But if the spectroscopic factors are renormalized to $\sigma=4$, the results become more or less compatible with each other. Making this renormalization, we overestimate the spectroscopic factors. We suggest, however, that the contribution of higher $^8$Li states is taken into account effectively to some extent in this way. 

We present mainly the results obtained using the spectroscopic factors from \cite{Kan}. Two versions (labeled as ``with CN'' and ``no CN'' in \cite{Kan}) give results very close to each other. We use below the former version. Also some examples for other sets are shown. The signs of the coefficients of fractional parentage are chosen according to the shell model calculation \cite{Kwas}, where the angular momentum addition scheme is exhibited explicitly, for all the sets.

The single particle wave functions of the proton, neutron, and $\Lambda$ hyperon needed for calculation of $S^i(r)$ in (\ref{e5}) are generated by a standard Woods-Saxon potential
\begin{equation}
V(r)=\frac{v_0}{1+\exp\left(\frac{r-R}{a}\right)}.\label{e9}
\end{equation}
Parameters $v_0=-73.9$ MeV, $R=2.5$ fm, and $a=0.65$ fm for the proton, and $v_0=-45.3$ MeV, $R=2.5$ fm, and $a=0.52$ fm for the neutron are fitted to provide the known separation energies\footnote{The parameters of the neutron potential are taken from \cite{Bert}.}.
For the $\Lambda$ hyperon, we take $v_0=-30$ MeV, $R=2.42$ fm, and $a=0.6$ fm, which gives $\Lambda$ binding energy $B_\Lambda=9.71$ MeV and, thus, $e_\Sigma=-77.36$ MeV. The Coulomb potential of the uniformly charged sphere is included for the proton and $\Sigma^-$ hyperon.

Quantitative knowledge of $\Sigma$-nucleus interaction is not yet achieved. The data obtained from the $(\pi^-,K^+)$ reaction in the kinematical region corresponding to production of a real $\Sigma^-$ hyperon \cite{Sah1} show that the $\Sigma^-$-nucleus interaction is repulsive in a wide range of $A$. On the other hand, the data on $\Sigma^-$ atoms exhibit attractive shifts (see \cite{BFG} and references therein). This seeming contradiction can be reconciled if the potential is mainly repulsive but has an attractive pocket in the peripherical region most important for atomic states \cite{BFG,HH,FG}. Shape of the potential cannot be fixed from existing data \cite{HH}. We choose the $\Sigma^--{}^9$Be potential according to \cite{Koh} also in form (\ref{e9}) with $R=2.59$ fm and $a=0.6$ fm. In most of calculations, we use $v_0=+30$ MeV.

\begin{table}
\caption{\label{tab1}$\Sigma^-$ admixture probabilities $p_{\Sigma^-}$ in units of 10$^{-4}$ for various coupling potentials. The repulsive $\Sigma^-$-nucleus potential ($v_0=+30$ MeV) and spectroscopic factors from \protect\cite{Kan} were used.}
\begin{ruledtabular}
\begin{tabular}{lcc}
Coupling & \multicolumn{2}{c}{$p_{\Sigma^-}$}\\
potential & $1^-$ & $2^-$\\
\hline
ESC08a & 0.57 & 0.98 \\
ESC08b & 1.90 & 2.12 \\
ESC08a$^{\prime\prime}$ & 6.74 & 5.00 \\
NHCD & 4.54 & 1.16 \\
NSC89 & 5.73 & 3.56 \\
D2 & 2.42 & 2.07
\end{tabular}
\end{ruledtabular}
\end{table}

Probabilities of $\Sigma^-$ admixture in the $1^-$ and $2^-$ states of $^{10}_\Lambda$Li are presented in Table \ref{tab1} for various coupling potentials. The spectroscopic factors are taken from \cite{Kan}. It is seen that the probabilities differ greatly for different coupling potentials. The modern models of the ESC type give the probabilities different by one order of magnitude. The lowest probabilities are obtained with coupling potential ESC08a while the highest ones -- with ESC08a$^{\prime\prime}$. The ESC08a$^{\prime\prime}$ potential has an intermediate-range component, which is much greater than those in the other cases. The ESC08b model as well as the older potentials gives results in between those two cases. This means, of course, that the probabilities cannot be predicted reliably now. But it is seen that the probabilities are always less than 0.1\%. 

\begin{table}
\caption{\label{tab2}The same as in Table \protect\ref{tab1} for the ESC08b coupling potential and various sets of spectroscopic factors.}
\begin{ruledtabular}
\begin{tabular}{lcc}
Spectroscopic & \multicolumn{2}{c}{$p_{\Sigma^-}$}\\
factors & $1^-$ & $2^-$\\
\hline
\protect\cite{Kan}, CN & 1.90 & 2.12 \\
\protect\cite{Jep}, T1 & 1.09 & 3.14 \\
\protect\cite{Jep}, T2 & 2.03 & 2.58 \\
\protect\cite{Kan}, VMC & 0.96 & 4.43 \\
\protect\cite{CK}, CK & 1.38 & 2.08 \\
\protect\cite{Tim}, STA & 1.83 & 2.50
\end{tabular}
\end{ruledtabular}
\end{table}

In Table \ref{tab2}, we show the same probabilities obtained with the coupling potential ESC08b and various sets of the spectroscopic factors. The T1 and T2 versions in \cite{Jep} correspond to different optical potentials in the final state used for analysis of the $d({}^9\mbox{Li},t){}^8$Li reaction. CK means the famous Cohen-Kurath shell model \cite{CK} while VMC refers to a variational Monte Carlo calculation \cite{Wir} (the corresponding spectroscopic factors are quoted in \cite{Kan}). Last, the set labeled as STA is taken from compilation \cite{Tim}.

It is seen that the probabilities depend on the choice of the spectroscopic factors significantly. However, this uncertainty is not so drastic as for the coupling potentials. Remind that the spectroscopic factors in all the cases are renormalized as described above.

All the probabilities in Tables \ref{tab1} and \ref{tab2} are obtained with the purely repulsive $\Sigma^--{}^9$Be potential ($v_0=+30$ MeV). We tried to add a small peripherical attractive pocket to this potential. The probabilities changed very slightly. Therefore, $\Sigma^-$ peripherical attraction in this hypernucleus is of small importance.

\begin{figure}
\includegraphics*[width=85mm]{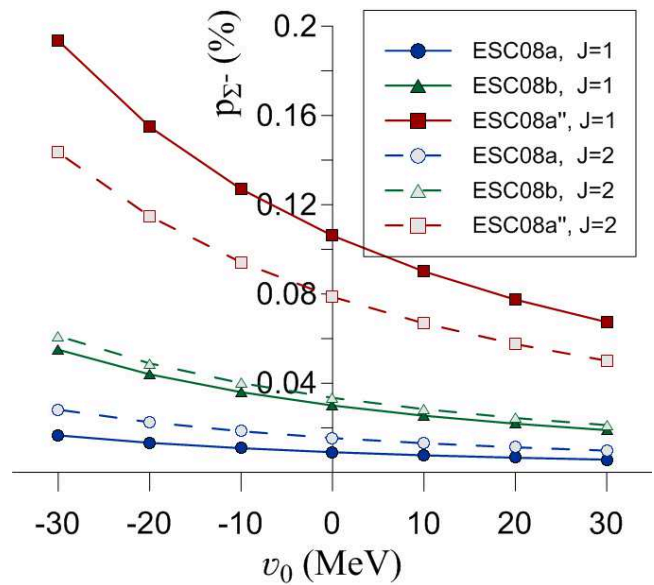}
\caption{\label{f1} (Color online) $\Sigma^-$ admixture probability dependence on the strength of the diagonal $\Sigma^-$-nucleus potential for the ESC coupling potentials.}
\end{figure} 

In Fig. \ref{f1}, we display the $p_{\Sigma^-}$ dependence on $v_0$ for the ESC coupling potentials. When the diagonal potential becomes attractive, the probabilities grow by several times. Only for the strongest coupling (ESC08a$^{\prime\prime}$) and $v_0=-(20\div 30)$ MeV, $p_{\Sigma^-}(1^-)$ is compatible with 0.18\% reported by Umeya and Harada \cite{Har1}.

\begin{figure}
\includegraphics*[width=85mm]{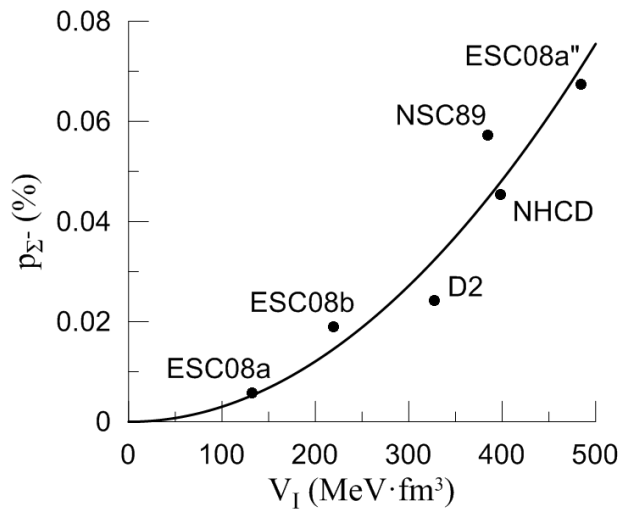}
\caption{\label{f2} $\Sigma^-$ admixture probability in the $1^-$ state as a function of volume integral $V_I=\int V_0(r)d^3r$ of the spin-independent coupling potential. The parabolic curve represents the best fit.}
\end{figure}

Fig. \ref{f2} shows $p_{\Sigma^-}(1^-)$ versus volume integral $V_I=\int V_0(r)d^3r$ of the spin-independent coupling potential. It is seen that $p_{\Sigma^-}(1^-)$ is determined mainly by spin-independent coupling potential $V_0$. The dependence on $V_I$ is nearly quadratic, which is natural in terms of the perturbation theory. It is interesting that we could not find a similar simple dependence of $p_{\Sigma^-}(2^-)$ on any component (or some combination of them) of coupling potential (\ref{e8}), though, of course, stronger coupling potentials lead to greater probabilities. 

\begin{figure}
\includegraphics*[width=85mm]{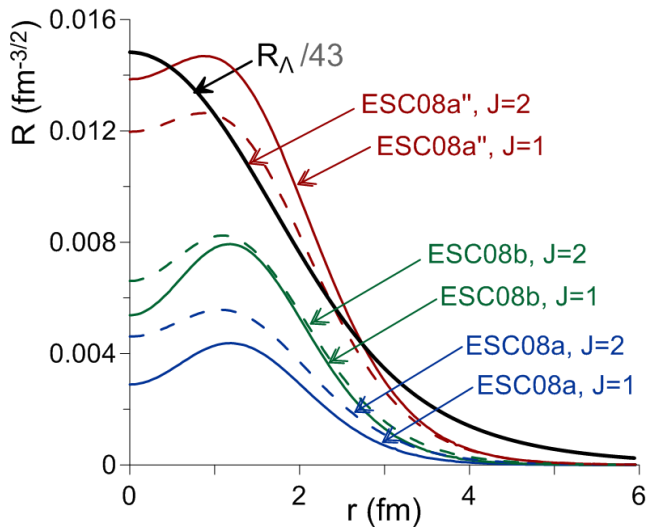}
\caption{\label{f3} (Color online) Radial wave functions of the $\Sigma^-$ component for the ESC coupling potentials in comparison with the scaled $\Lambda$ wave function. Solid (dashed) curves are for the $1^-$ ($2^-$) state.}
\end{figure}

Finally, we demonstrate wave functions $R_\Sigma(r)=u_\Sigma(r)/r$ for the ESC models in Fig. \ref{f3}. Shape of the wave function (shown in comparison with the $1s$ $\Lambda$ wave function) clearly reveals its distinction from usual bound state $1s$ wave functions. $R_\Sigma$ has a minimum at $r=0$, which is some reminiscence that it can be considered as a superposition of continuum states. We checked that Coulomb $\Sigma^-$-nucleus interaction plays a minor role in this problem so atomic bound states contribute little to this superposition. Maybe, the situation is different in heavier systems. On the other hand, the wave function drops abruptly at the asymptotics due to the large absolute value of $e_\Sigma$. Therefore, the $\Sigma^-$ distribution is concentrated near the nuclear surface as already was stressed in \cite{TL01}.

\section{\label{s4}Conclusion and outlook}
In this paper, we revisit the problem of $\Sigma^-$ admixture in the $^{10}_\Lambda$Li hypernucleus, which is important for understanding of reactions producing neutron-rich $\Lambda$ hypernuclei. The simple approach incorporating some features of the shell model enables us to deal with both attractive and repulsive $\Sigma$-nucleus potentials. We show that $\Sigma$-nucleus repulsion reduces the $\Sigma$ admixture probabilities by several times with respect to attractive potentials.

We cannot predict the $\Sigma^-$ admixture probabilities since the coupling potential is extremely uncertain. However, our calculations for very different coupling potentials give always $p_{\Sigma^-}<0.1$\%, much less than those (about 0.5\%, \cite{HUH}) needed to explain the cross sections of the $^{10}\mbox{B}(\pi^-,K^+){}^{10}_\Lambda$Li by the one-step $\pi^-p\to K^+\Sigma^-$ mechanism. Though our model involves some approximations it is unlikely that they can lead to so large underestimation of the probability. Possibly, the so-called ``coherent mixing'' \cite{Aka} giving large $\Sigma$ probabilities in $^4_\Lambda$H and $^4_\Lambda$He is especially effective when all the baryons are in the same ($1s$) state while it is less efficient in $p$ shell systems.

Therefore, the conclusion that the one-step $\pi^-p\to K^+\Sigma^-$ mechanism is dominant for the $(\pi^-,K^+)$ reaction \cite{HUH} seems to be questionable and further investigations are needed. Possibly, the structure of the $\Sigma^-$ component is more complicated. Our tentative estimation indicates that the probability of the $\Sigma^-$ hyperon $d$ state may be not so small. The two-step mechanism with meson charge exchange is also not understood completely. The models applied in \cite{TL03,HUH} are rather simplified while it is known that dynamics of the similar process, namely, pion double charge exchange, is much more complex. Last, maybe another one-step mechanism, i.e., production of neutron-rich $\Lambda$ hypernuclei via $\Delta$ admixture in a target nucleus, e.g., $\pi^-\Delta^{++}\to K^+\Lambda$, suggested in \cite{L04} deserves attention.
\begin{acknowledgments}
This work was supported in part by Russian Foundation for Basic Research, grant No. 12-02-01045.
\end{acknowledgments}

\end{document}